# Detection of Explosives by Using a Neutron Source Based on a Proton Linac

*S. N. Dolya*

*Joint Institute for Nuclear Research, Joliot - Curie 6, Dubna, Russia, 141980*

**Abstract**

The paper considers an opportunity of detecting explosives by using radiation capture of a neutron with nitrogen nucleus $N^{14}$ (n, γ) $N^{15}$. Proton LINAC is offered as the neutron source with the following parameters: proton energy $Є$ = 5 MeV, beam pulse current $I_{pulse}$ = 1.7 mA, duration of the current pulse $\tau_{pulse}$ = 200 μs, repetition rate F = 50 Hz. The reaction in which neutrons are formed is $Li^7$ (p, n) $Be^7$. It is shown that this neutron source will have the intensity of $10^{11}$ n / s that will allow one to detect explosives of the size of a tennis ball.

**1. Introduction**

As it is shown by in chemical formulae [1], explosive chemicals have increased nitrogen content. This nitrogen can be determined by the neutron capture $N^{14}$ (n, γ) $N^{15}$ and results in γ – quanta production with the energy of 3-5 MeV, [2], p. 884.

**2. The desired intensity of the neutron flux**

We calculate the conditions of detection of explosives having the size of a tennis ball. The cross section of the reaction $N^{14}$ (n, γ) $N^{15}$, measured at the reactor neutron beam [3], is equal to: $\sigma_{N14\ (n,\ \gamma)\ N15}$ = 8 * $10^{-26}$ $cm^2$.

We find the density of nitrogen atoms in hexogen from the following considerations. The molecular weight of hexogen is equal to 227 g / mol, density $\rho_{RDX}$ = 1.77 g / $cm^3$. To find the density of molecules in hexogen, we compose the following proportion:

$$6*10^{23} \text{ ---------------- } 227 \text{ g}$$
$$x \text{ -------------- } 1.77 \text{ g,}$$

where x = 4.7 * $10^{21}$. Nitrogen in the hexogen gram - molecule contains 84 g, the mass ratio of nitrogen to the molecular mass of RDX is: 84/237 = 0.37. If this value multiplied by the density of the obtained molecules in hexogen, we will get the density of nitrogen atoms $n_N$ = 1.7 * $10^{21}$ atoms/cubic centimeter.

The probability of the radiation capture of the neutron in hexogen $p_1$ is equal to the following:

$$p_1 = \sigma n l, \quad (1)$$

where $\sigma_{N14\,(n,\,\gamma)\,N15} = 8 * 10^{-26}$ cm$^2$, the radiation capture cross section for $n = 1.7 * 10^{21}$ which is the density of nitrogen atoms in hexogen, $l \approx 3$ cm is the length of the interaction of neutrons with hexogen atoms. It is approximately equal to the radius of a tennis ball that is equal to 3 cm. Substituting the numbers into formula (1) we find that the probability of the interaction of the neutron with the nitrogen atoms in the volume of the explosive is equal to the following: $p_1 = \sigma n l = 4 * 10^{-5}$.

Let us find geometrical factor $G_1$ equal to the ratio of the cross-section of the explosive $S_{ex}$ to the total volume of neutron spreading out: $G_1 = S_{ex} / 4\pi r_1^2$, where $r_1$ is the distance from the target generating the neutron flux till the object under study. Considering the cross-section of the explosive to be equal to $S_{ex} = 30$ cm$^2$ and the distance from the target to the object under study $r_1 = 3$ m, we will find that the value $G_1 = S_{ex} / 4\pi r_1^2 = 3 * 10^{-5}$.

Gamma quanta emitted in the reaction $N^{14}$ (n, $\gamma$) $N^{15}$ are distributed isotropic. We assume that the radius of the sphere $r_2$ where the detectors are located is equal to $r_2 = 1$ m and the area of the detectors of gamma quanta $S_{det}$ is equal to $S_{det} = 1$ m$^2$. Then we find that the geometric factor $G_2$ equal to the ratio of the detector area to the total area around the object under study ($S_t = 4\pi r_2^2 = 10$ m$^2$), is $G_2 = S_{det} / S_t = 10^{-1}$.

We assume the probability of detecting gamma quanta with the energy of 3-5 MeV to be equal to 1. Then, the probability $p_2$ of registering the gamma quant from the reaction $N^{14}$ (n, $\gamma$) $N^{15}$ per one neutron, is equal to the following:

$$p_2 = p_1 * G_1 * G_2 = 4*10^{-5}*3*10^{-5}*10^{-1} = 10^{-10}. \quad (2)$$

Equation (2) shows that, in order to register one gamma quant per second, it is required to have the neutron flux equal to $N_n = 10^{10}$ n / s. To reliably detect the obtained gamma quanta, it is required to have the neutron flux by one order of magnitude greater than the above: $N_n = 10^{11}$ n / s.

**3. Linear proton accelerator as a neutron source.**

We choose the nuclear reaction leading to the neutron production: $Li^7$ (p, n) $Be^7$, [4]. The chosen energy of the accelerated protons is equal to $\varepsilon_p$ = 5 MeV. For this proton energy the neutron yield from a thick lithium target can be estimated as $Y = 10^{-3}$. This means that the number of accelerated protons per second must be equal to $N_p = 10^{14}$ p / s or proton beam current must be equal to $I_p$ = 17 µA. The target for the proton beam current $I_p$ = 2 mA has been considered in [5].

Linear accelerators usually operate in the pulsed mode: after a short pulse there is quite a long pause. So, for the high-current linear proton accelerator considered in detail in [6], the pulse duration $\tau_{pulse}$ is $\tau_{pulse}$ = 200 µs, the pulse repetition rate F = 5 Hz. Thus, the time of the accelerator operation is a small value in comparison with the real period of time: $\tau_{pulse} * F = 10^{-3}$. This means that for the required proton current $I_p$ = 17 µA and the pulsed current should be of the value of $I_{p\ pulse}$ = 17 mA.

It will be shown that this value of the pulsed current will cause a serious problem of holding the longitudinal and transverse dimensions of the beam. Therefore, we choose the value of the pulsed current to be equal to $I_{p\ pulse1}$ = 1.7 mA. To maintain the same average current value, we choose the pulse repetition rate F = 50 Hz.

As a proton linac accelerating structure we use a spiral waveguide. A spiral waveguide is a conventional coaxial cable whose central conductor is a spiral. The acceleration of the intense (100 mA) proton beam in a spiral waveguide was considered in [6]. We take some considerations from this work. The spiral waveguide allows one to accelerate the high intensity proton beams (100 mA) at the high acceleration rate (1 MeV / m).

In the longitudinal direction the proton bunches are kept by the field of the accelerating wave. In the transverse (radial) direction the accelerating wave defocuses proton bunches which is added with the defocusing caused by the forces of Coulomb repulsion.

*3. 1. Stability of longitudinal oscillations of protons*

We choose the injection voltage U = 800 kV. This means that the initial velocity of the protons in the beam is equal $\beta_{in} = 4 * 10^{-2}$, where $\beta = v / c$ which is the velocity expressed in units of the velocity of light in vacuum: $c = 3*10^{10}$ cm / s. We assume that the acceleration is performed at frequency f = 300 MHz [6], the radius of the beam is assumed to be $r_b$ = 1 mm.

The number of particles in each bunch $N = I_p * 6 * 10^{18} / f \approx 3.4 * 10^7$ p / bunch. The slowed down wave length at the beginning of the acceleration is equal to $\lambda_s = \lambda * \beta_{in} = 4$ cm, where $\lambda = c / f = 1$ m is the wavelength in vacuum. The bunch in the bunched beam occupies the value over the phases approximately equal to $\pm 18^0$, so that the length of the proton bunch in the longitudinal direction at the beginning of the acceleration is equal to $l_b$ = 4 mm. The volume occupied by the bunch, is equal to $V_p = \pi r_b^2 * l_b = 1.26 * 10^{-2}$ cm$^3$, and the density of protons in the bunch for the beam current $I_{pulse}$ = 1.7 mA, is equal to $n_p = 2.7 * 10^9$.

Plasma frequency $\omega_p$ is equal to $(4\pi e^2 n_p / m)^{1/2} = 1.3 * 10^3 * n_p^{1/2} = 6.7 * 10^7$. If you do not use any power to maintain the size of the bunch in the longitudinal direction, the bunch will spread out in the longitudinal direction according to the following formula: $l = l_{in} * \exp[f_p t]$, where we have introduced $f_p = \omega_p / 2\pi$. In the longitudinal direction while accelerating the beam the sizes of the bunch are kept by the wave field accelerating the bunch.

Now we choose the amplitude value of the accelerating field to be equal to $E_0$ = 10 kV / cm, and the cosine of the synchronous phase $\cos \varphi_s$ = 0.5. Then the specific acceleration equal to the ratio of the increase of the proton energy at the wavelength to the proton rest energy $mc^2$ = 1 GeV, [6], is equal to the following:

$$W_\lambda = eE\lambda*\cos\varphi_s / mc^2 = 5*10^{-4}. \qquad (3)$$

The phase oscillation frequency $\Omega_{ph}$ is the frequency at which the protons in the bunch oscillate in the longitudinal direction:

$$\Omega_{ph} = \omega*(W_\lambda * tg\, \varphi_s / 2\pi\beta_s)^{1/2} = 1.2*10^8. \qquad (4)$$

This frequency $\Omega_{ph} = 1.2 * 10^8$ is more than the plasma frequency $\omega_p = 6.7 * 10^7$. This shows that the bunch will be kept in the longitudinal direction by the wave field.

*3. 2. Stability of transverse oscillations of protons*

A spiral waveguide has very small transverse dimensions: the outer diameter size of the screen is smaller than 10 cm. Then it is possible to withhold the proton beam in the transverse direction by means of the longitudinal solenoid magnetic field. When you have a great acceleration energy rate and high accelerated beam intensity, it is required to have a big magnetic field of 10 T to hold this beam in the transverse direction [6]. This magnetic field can be obtained by using superconducting coils, but it is inconvenient while the accelerator is operating under almost field conditions.

Let us consider focusing of the proton beam by means of magnetic quadrupole lenses made on the basis of the permanent magnets of the NdFeB. The focusing of the intense proton beams by using the quadropole lenses has been considered in detail in [7].

The length of the period of the magnetic quadrupoles along the z-axis is denoted by value l. Then, in the transverse directions (x, y) the protons will be rejected with frequency $\omega = v / l$, where v is the longitudinal velocity of protons. Let us choose l to be equal to: $l = 10$ cm.

To have the same coefficients in the equation of the transverse motion over all the length of the accelerator, it is possible to increase the length of the period of the focusing system in accordance with the increase of the velocity. We remind that in this case the rate will increase by 2.5 times: from $\beta_{in} = 4 * 10^{-2}$ to $\beta_{fin} = 10^{-1}$, that corresponds to the finite proton energy $\mathcal{E} = 5$ MeV.

The frequency of the transverse oscillations determined by the focusing force acting on the protons from the side of the quadrupole lenses, in dimensionless variables is equal to [7] $evG / mc\omega^2$. The equations of motion on the transverse (x, y) coordinates can be written as follows:

$$d^2 x/d\tau^2 - [(f_p^2 + f_r^2)/\omega^2 + (eGl^2/mcv)\cos(2\tau)]x = 0, \qquad (5)$$

where we have introduced the following: $\tau = vt / l$, $\Omega_r = \Omega_{ph} / 1.41$, $f_r = \Omega_r / 2\pi$ – is the growth rate of the transverse deflection of protons under the action of the accelerating electric field, $G = dH_x / dx$, $dH_y / dy$ is the transverse magnetic field gradient.

Equation (5) is the equation of Mathieu:

$$d^2x/d\tau^2 - [a + 2q*\cos(2\tau)]x = 0, \qquad (6)$$

where a = $(f_p^2 + f_r^2) / \omega^2$, q = $eGl^2 / 2mcv$. Depending on the values of coefficients a and q, the solutions may be stable or unstable. The diagram of the stability region for the most exciting case of a > 0, q < 1 is given in [8]. For the chosen parameters $f_p = 1.1 * 10^7$, $f_r = 1.35 * 10^7$, $\omega = v / l = 1.2 * 10^8$, we find that a = $(f_p^2 + f_r^2) / \omega^2 = 3 * 10^{14}/1.44*10^{16} = 0.02$. We choose the magnetic field gradient in the lenses to be equal to G = 1.7 kGs / cm. Then
q = $eGl^2 /2mcv = 5*10^{-10}*1.7*10^3*10^2/2*1.7*10^{-24}*3*10^{10}*1.2*10^9 = 0.7$.
The point with coordinates a = 0.02 and q = 0.7 gets into the middle of the region of simultaneous stability on the values of a and q, [8]. It means that in this case we will have simultaneous stability of deviations in the directions x and y.

**4. The required high-frequency power**

We find the high-frequency power required to achieve field strength $E_0 = 10$ kV / cm from the following relation [6]:

$$P = (c/8) E_0^2 r_0^2 [ kk_3/k_1^2 ]\{(1+I_0K_1/I_1K_0)(I_1^2-I_0I_2) +$$

$$+ \varepsilon (I_0/K_0)^2(1+I_1K_0/I_0K_1)(K_0K_2-K_1^2)\}, \quad (7)$$

where k = $\omega / c$ is the wave vector, $k_1 = k (1 / \beta_{ph}^2 - 1)^{1/2}$ –the radial wave vector inside the spiral, $k_3 = \omega / v_{ph}$ is the wave vector in the axial direction; $I_0$, $I_1$, $I_2$ are the modified function of Bessel of the first kind, $K_0$, $K_1$, $K_2$ – the modified Bessel functions of the second kind. The first term in the curly brackets corresponds to the power flux propagating inside the spiral and the second term corresponds to the power flux propagating between the spiral and screen. If this area is filled with dielectric having dielectric constant ε, then before the second term there must be factor ε.

In the case of significant wave slowing down ($\beta_{ph} << 1$) and of the optimum ratio between the perimeter of the spiral turn $2\pi r_0$ and delayed wavelength $\lambda_s = \beta_{ph} * \lambda$, the formula (7) is simplified. The expression $[kk_3 / k_1^2]$ can be replaced by $\beta_{ph}$. The value of the argument of the Bessel functions: x = 1, is equal to the value of { } in formula (7), i.e., { } = 4.44.

Let us assume that for the beginning of acceleration the value in the curved brackets in formula (7) is equal to { } = 8. We write formula (7) in the simplified form as follows:

$$P = (c/8) E_0^2 r_0^2 \beta_{in}\{\}. \quad (8)$$

Substituting the numbers into formula (8), we obtain the following value:

$$P = 3*10^{10}*10^{8}*4*10^{-2}*8/(8*9*10^{4}*10^{7}) = 133 \text{ kW},$$

where we have taken into account that 300 V / cm = 1 CGS unit of the electric field strength as well as 1 W = $10^7$ erg / s.

This high-frequency power can be obtained from a small-sized high-frequency generator. The length of the accelerator for the finite energy $\mathcal{E}$ = 5 MeV at the rate of the energy gain $eE_0 \cos \varphi_s$ = 0.5 MeV / m, will be equal to $L_{acc}$ = 10 m.

Below there is a table of parameters of the accelerator.

Parameters of the accelerator

| Option | Value |
| --- | --- |
| The voltage of the proton source is, kV | 800 |
| The initial radius of the spiral $r_{0in}$, cm | 1 |
| Beam radius $r_b$, cm | 0.1 |
| Bunched beam current, mA | 1.7 |
| The frequency of the acceleration f, MHz | 300 |
| The gradient of the magnetic field, kGs/cm | 1.7 |
| The duration of the current pulse $\tau_b$, μs | 200 |
| The tension of electric field $E_0$, MV /m | 1 |
| The cosine of the synchronous phase, $\cos\varphi_s$ | 0.5 |
| The pulse repetition rate F, Hz | 50 |
| High-frequency power, kW | 133 |
| The length of the accelerator, m | 10 |

**5. Conclusion**

The use of the neutron source based on a proton linear accelerator to search for explosives allows one to confidently detect the explosive substance having dimensions of the order of a tennis ball. Since the focusing of the proton beam in such an accelerator is carried out by means of permanent magnets, it is required to have only a high-frequency generator to feed this accelerator.

The neutron source can be made on the basis of photonuclear reactions. In particular, it is possible to produce quite enough neutrons by means of the microtron [9]. However, the microtron also produces the intensive flux of gamma quanta against which we have to undertake appropriate measures of radiation protection.

References

1. http://ru.wikipedia.org/wiki/Тринитротолуол
   http://ru.wikipedia.org/wiki/Гексоген

2. Tables of physical quantities, Handbook ed. I. K. Kikoin, Moscow, Atomizdat, 1976

3. A.I. Egorov, R.I. Krutov, Y.E. Loginov, S.E. Malyutenkova, Measurement of cross sections for radiative capture of thermal neutrons by nuclei $^{14}$N and $^{19}$F by γ- spectroscopy neutron beam reactor, Preprint St. Petersburg Institute of Nuclear Physics, them. B.P. Konstantinov, Gatchina, 2004

4. J. H. Gibbons and R. L. Macklin, Total Neutron Yields from Light Elements under Proton and Alpha Bombardment, Physical Review, v. 114, number 2, p. 571, 1959

5. B. N. Lee, J. A. Park, Y. S. Lee, e. a, Design of Neutron Target with the 4 MeV Cyclotron for BNCT, Journal of the Korean Physical Society, Vol. 59, No. 2, August 2011, pp. 2032-2034

6. S. N. Dolya, A multy beam proton accelerator, http://arxiv.org/ftp/arxiv/papers/1509/1509.04158.pdf

7. I. M. Kapchinsky, Particle dynamics in linear resonance accelerators, Moscow, Atomizdat, 1966

8. V. Paul, electromagnetic traps for charged and neutral particles, Nobel Lecture, 1989, UFN, t.160 in. 12, http://ufn.ru/ufn90/ufn90_12/Russian/r9012d.pdf

9. S. N. Dolya, Microtron for Smog Particle Photo Ionization, https://www.researchgate.net/publication/283536709_Microtron_for_Smog_Particles_Photo_Ionization